\documentclass{article}
\usepackage{slashed,feynmf,epsfig,amsmath,amssymb,enumitem}
\usepackage[papersize={8.5in,11in}]{geometry}
\renewcommand{\vec}[1]{\boldsymbol{\mathrm{#1}}}
\usepackage{graphicx}
\begin{document}
\title{Structure Growth and the CMB in Modified Gravity (MOG)}
\author{J. W. Moffat\\~\\
Perimeter Institute for Theoretical Physics, Waterloo, Ontario N2L 2Y5, Canada\\
and\\
Department of Physics and Astronomy, University of Waterloo, Waterloo,\\
Ontario N2L 3G1, Canada}
\maketitle




\begin{abstract}
An important piece of evidence for dark matter is the need to explain the growth of structure from the time of horizon entry and radiation-matter equality to the formation of stars and galaxies. This cannot be explained by using general relativity without dark matter. So far, dark matter particles have not been detected in laboratory measurements or at the LHC. We demonstrate that enhanced structure growth can happen in a modified gravity theory (MOG). The vector field and particle introduced in the theory to explain galaxy and cluster dynamics plays an important role in generating the required structure growth. The particle called the phion (a light hidden photon) is neutral and is a dominant, pressureless component in the MOG Friedmann equations, before the time of decoupling. The dominant energy density of the phion particle in the early universe, generates an explanation for the growth of density perturbations. The angular acoustical power spectrum due to baryon-photon pressure waves is in agreement with the Planck 2013 data. As the universe expands and large scale structures are formed, the density of baryons dominates and the rotation curves of galaxies and the dynamics of clusters are explained in MOG, when the phion particle in the present universe is ultra-light. The matter power spectrum determined by the theory is in agreement with current galaxy redshift surveys.
\end{abstract}

\maketitle

\section{Introduction}

Dark matter was introduced to explain the stable dynamics of galaxies and galaxy clusters. General relativity (GR) with only ordinary baryon matter cannot explain the present accumulation of astrophysical and cosmological data without dark matter. However, dark matter has not been observed in laboratory experiments~\cite{Wimp,Xenon,Lux}. Therefore, it is important to consider a modified gravitational theory. The observed acceleration of the universe has also complicated the situation by needing a dark energy, either in the form of the cosmological constant vacuum energy or as a modification of GR.

The fully relativistic and covariant modified gravitational theory (MOG) Scalar-Tensor-Vector-Gravity (STVG)~\cite{Moffat,BrownsteinMoffat1,BrownsteinMoffat2,BrownsteinMoffat3,Brownstein,MoffatToth1,MoffatToth2,MoffatToth3,MoffatToth4,MoffatRahvar1,MoffatRahvar2,Roshan1,Roshan2} has been successfully applied to explain the rotation curves of galaxies and the dynamics of galaxy clusters. In addition to the attractive gravitational force proportional to the gravitational constant $G$, the theory also has a repulsive gravitational force generated by a massive Proca vector field $\phi_{\mu}$.

A modified gravity theory based only on baryonic matter cannot explain the growth of matter, before the onset of decoupling at the surface of last scattering about 380,000 years after the big bang. The baryons couple to the photons producing pressure that prevents the overdense instability needed for growth structure. The damping and the dissipative pressure behavior of baryons coupled to photons prevent a GR model without non-baryonic dark from fitting the angular acoustical power spectrum. The MOG cannot explain the growth of structure to form galaxies and galaxy clusters assuming only the dominance of baryonic matter before decoupling.

In the following, we will extend the explanation for the growth of structure and the CMB data proposed originally in STVG~\cite{Moffat}. We will show that when the density of the neutral, pressureless vector particle called the phion particle, associated with the field $\phi_{\mu}$, dominates in the early universe before the time of decoupling, MOG can lead to a mechanism for the growth of perturbations faster than GR, solving the cosmological structure problem. The early time domination of the $\rho_\phi$ density and the later time domination of the baryon density $\rho_b$, together with the later time enhancement of the gravitational constant $G$, determine 
the acoustical power spectrum in agreement with the Planck 2013 data~\cite{Planck}. The theory also determines the matter power spectrum in agreement with current large scale galaxy redshift surveys.

The mass of the phion is determined by a scalar field $\mu(x)$. The value of $\mu$ that fits the galaxy rotation curves and the cluster dynamics is $\mu=0.04\,{\rm kpc}^{-1}$~\cite{MoffatRahvar1,MoffatRahvar2}, corresponding to a mass $m_\phi=2.6\times 10^{-28}\,{\rm eV}$. The mass of the phion in the early universe, before decoupling (surface of last scattering) is heavier, $m_\phi\gg 10^{-28}\,{eV}$, so the phion behaves like cold dark matter (CDM), and its mass decreases as the universe expands becoming the ultra-light phion particle today. 

\section{MOG Field Equations}

The MOG theory has a fully covariant action composed of scalar, vector and tensor fields~\cite{Moffat}:
\begin{equation}
\label{action1}
S=S_G+S_\phi+S_S+S_M.
\end{equation}
The components of the action are the Einstein gravity action ($c=1)$:
\begin{equation}
S_G=\frac{1}{16\pi}\int d^4x\sqrt{-g}\Big[\frac{1}{G}(R+2\Lambda)\Big],
\end{equation}
the massive vector field $\phi_\mu$ action:
\begin{eqnarray}
S_\phi= - \omega\int d^4x\sqrt{-g}\Big[\frac{1}{4}B^{\mu\nu}B_{\mu\nu}-V(\phi)\Big],
\end{eqnarray}
and the action for the scalar fields $G$ and $\mu$:
\begin{eqnarray}
S_S=\int d^4x\sqrt{-g}\Big[\frac{1}{G^3}\Big(\frac{1}{2}g^{\mu\nu}\nabla_\mu G\nabla_\nu G-V(G)\Big)
+\frac{1}{\mu^2 G}\Big(\frac{1}{2}g^{\mu\nu}\nabla_\mu\mu\nabla_\nu\mu - V(\mu)\Big)\Big].
\label{scalar}
\end{eqnarray}
Here, $\nabla_\mu$ denotes the covariant derivative with respect to the metric $g_{\mu\nu}$, $\omega$ is a dimensionless coupling constant and $V(\phi), V(G)$ and $V(\mu)$ denote potentials for the fields $\phi_\mu,
G$ and $\mu$, respectively.  In particular, we have
\begin{equation}
\label{Vphi}
V(\phi)=-\frac{1}{2}\mu^2g^{\mu\nu}\phi_\mu\phi_\nu + W(\phi),
\end{equation}
where $W(\phi)$ is a self-interaction potential.

The energy-momentum tensor is defined as
\begin{equation}
T_{\mu\nu}=T_{M\mu\nu}+T_{\phi\mu\nu}+T_{S\mu\nu},
\end{equation}
where
\begin{equation}
T_{X\mu\nu}=-\frac{2}{\sqrt{-g}}\frac{\delta S_X}{\delta g^{\mu\nu}},\quad (X=[G,\phi,S,M]).
\end{equation}
We have
\begin{equation}
\label{Tphi}
T_{\phi\mu\nu}=\omega\biggl({B_\mu}^\alpha B_{\alpha\nu}-\frac{1}{4}g_{\mu\nu}B^{\alpha\beta}B_{\alpha\beta}+g_{\mu\nu}V(\phi)-2\frac{\partial V(\phi)}{\partial g^{\mu\nu}}\biggr).
\end{equation}

A variation of the action with respect to $g^{\mu\nu}$ yields the field equations:
\begin{equation}
G_{\mu\nu}-g_{\mu\nu}\Lambda+Q_{\mu\nu}=8\pi GT_{\mu\nu},
\end{equation}
where
\begin{equation}
Q_{\mu\nu}=G\biggl(g^{\alpha\beta}\Theta\nabla_\alpha\nabla_\beta g_{\mu\nu}-\nabla_\mu\nabla_\nu\Theta\biggr),
\end{equation}
and where $\Theta(x)=G^{-1}(x)$.

\section{Friedmann Equations and Cosmology}

We base our cosmology on the homogeneous and isotropic Friedmann-Lemaitre-Robertson-Walker (FLRW) background metric:
\begin{equation}
ds^2=dt^2-a^2(t)\biggl(\frac{dr^2}{1-Kr^2}+r^2(d\theta^2+\sin^2\theta d\phi^2)\biggr),
\end{equation}
where $K=-1,0,1\,({\rm length}^{-2})$ for open, flat and closed universes, respectively. We use the energy-momentum tensor of a perfect fluid:
\begin{equation}
T_{\mu\nu}=(\rho+p)u^\mu u^\nu-pg_{\mu\nu},
\end{equation}
where $u^\mu=dx^\mu/ds$ and $\rho$ and $p$ are the density and pressure of matter. We have
\begin{equation}
\rho=\rho_M+\rho_\phi+\rho_G+\rho_\mu,
\end{equation}
where $\rho_M,\rho_\phi, \rho_G$ and $\rho_\mu$ denote the density of matter, the (neutral) vector field $\phi_\mu$ (phion particle) and the scalar fields $G$ and $\mu$, respectively.

Due to the symmetries of the FLRW background spacetime, we have $\phi_0\neq 0,\phi_i=0\,(i=1,2,3)$ and $B_{\mu\nu}=0$. The MOG Friedmann equations are given by:
\begin{equation}
\label{Friedmann1}
\biggl(\frac{\dot a}{a}\biggr)^2+\frac{K}{a^2}=\frac{8\pi G\rho}{3}+\frac{\dot a}{a}\frac{\dot G}{G}+\frac{\Lambda}{3},
\end{equation}
\begin{equation}
\label{Friedmann2}
\frac{\ddot a}{a}=-\frac{4\pi G}{3}(\rho+3p)+\frac{1}{2}\biggl(\frac{{\ddot G}}{G}-\frac{2{\dot G}^2}{G^2}+\frac{{\dot a}{\dot G}}{aG}\biggr)+\frac{\Lambda}{3}.
\end{equation}

In the following, we assume a spatially flat universe $K=0$. Furthermore, we assume that the time dependence
of the gravitational field strength is negligible, $\dot G\sim 0$, and we also assume that $\dot\mu\sim 0$. We obtain the approximate Friedmann equations:
\begin{equation}
\label{Friedmann3}
H^2=\frac{8\pi G\rho}{3}+\frac{\Lambda}{3},
\end{equation}
\begin{equation}
\frac{\ddot a}{a}=-\frac{4\pi G}{3}(\rho+3p)+\frac{\Lambda}{3},
\end{equation}
where $H=\dot a/a$. The energy conservation equation is
\begin{equation}
\dot\rho+3\frac{d\ln a}{dt}(\rho+p)=0.
\end{equation}
For the matter dominated universe with $p=0$ we have $\rho_M\propto 1/a^3$.

We introduce comoving coordinates:
\begin{equation}
{\bf x}=\frac{a_0{\bf r}}{a(t)}.
\end{equation}
The Fourier expanded density, pressure and gravitational potential in terms of plane waves with comoving wavenumber ${\bf k}$ are given by
\begin{equation}
\delta\rho({\bf r,t})=\frac{1}{(2\pi)^{3/2}}\bar\rho\int d^3k\delta_{\bf k}(t)\exp(i(a_0/a){\bf k}\cdot{\bf r}),
\end{equation}
\begin{equation}
\delta p({\bf r,t})=\frac{1}{(2\pi)^{3/2}}\int d^3k\delta p_{\bf k}(t)\exp(i(a_0/a){\bf k}\cdot{\bf r}),
\end{equation}
\begin{equation}
\Phi({\bf r,t})=\frac{1}{(2\pi)^{3/2}}\int d^3k\Phi_{\bf k}(t)\exp(i(a_0/a){\bf k}\cdot{\bf r}),
\end{equation}
where $\delta=\delta\rho/{\bar\rho}$ is the relative density perturbation contrast. We have for the gravitational potential:
\begin{equation}
k^2\Phi_{\bf k}=-4\pi G\biggl(\frac{a}{a_0}\biggr)^2\bar\rho\delta_{\bf k}.
\end{equation}

The modified Newtonian acceleration law in MOG for a point particle is given by
\begin{equation}
\label{MOGacceleration}
{\ddot r}=-\frac{G_\infty M}{r^2}\biggl[1-\frac{\alpha}{1+\alpha}(1+\mu r)\exp(-\mu r)\biggr],
\end{equation}
where $G_\infty=G_N(1+\alpha)$ and $G_N$ is Newton's constant.  This in turn defines the value of $\alpha$:
\begin{equation}
\label{alphaeq}
\alpha=(G_\infty-G_N)/G_N.
\end{equation}
The point mass acceleration law (\ref{MOGacceleration}) has been extended to continuous distributions of matter using the MOG Poisson
equation~\cite{BrownsteinMoffat1,MoffatToth4,MoffatRahvar1,MoffatRahvar2}. The MOG potential for a given density $\rho({\vec x})$ is
\begin{equation}
\Phi(\vec{x}) = - G_N \int d^3x'\frac{\rho(\vec{x}')}{|\vec{x}-\vec{x}'|}\Big[1+\alpha -\alpha
e^{-\mu|\vec{x}-\vec{x}'|}\Big]. 
\label{potential}
\end{equation}

\section{MOG Structure Growth}

In the present scenario there are five components to the energy density:
\begin{enumerate}
\item  neutral pressureless phion particles ($\phi_\mu$ field),

\item  baryonic matter,

\item  photons,

\item  neutrinos,

\item  Dark energy
\end{enumerate}

Thus, we have
\begin{equation}
\rho=\rho_m+\rho_r+\rho_d,
\end{equation}
where
\begin{equation}
\rho_m=\rho_b+\rho_\phi+\rho_G+\rho_\mu,\quad \rho_r=\rho_\gamma+\rho_\nu,
\end{equation}
and where $\rho_r, \rho_d, \rho_\gamma$ and $\rho_\nu$ denote the radiation, dark energy, photon and neutrino densities, respectively.

We assume that when $\rho_\phi$ dominates in the early universe before decoupling $\alpha < 1$. From (\ref{alphaeq}) we deduce that 
\begin{equation}
G\sim G_\infty\sim G_N.
\end{equation}
We anticipate that the parameter $\alpha$ will increase as the density of matter grows significantly near the big bang at $t=0$, so near $t=0$, $G=G_N(1+\alpha)$ will grow and increase the strength of gravitational attraction. The particle (field) densities are expressed as the ratios $\Omega_x=8\pi G_N\rho_x/3H^2$. In particular, we have for the baryon, phion particle and the cosmological constant $\Lambda$:
\begin{equation}
\label{Omegaeqs}
\Omega_b=\frac{8\pi G_N\rho_b}{3H^2},\quad\Omega_\phi=\frac{8\pi G_N\rho_\phi}{3H^2},\quad \Omega_{\Lambda}=\frac{\Lambda}{3H^2}.
\end{equation}
At the time of big-bang nucleosynthesis (BBN), we have $G\sim G_N$, guaranteeing that the production of elements at the time of BBN agrees with observations.  After decoupling $\rho_\phi\sim\rho_b$ until stellar and galaxy formation when $\rho_\phi\ll \rho_b$ and the MOG non-relativistic acceleration law sets in to explain the rotation curves of galaxies and the dynamics of clusters~\cite{Moffat,MoffatRahvar1,MoffatRahvar2}.

We assume that at horizon entry until some time after decoupling, $\rho_\phi\gg \rho_b$, $\rho_\phi\gg\rho_G$, $\rho_\phi\gg\rho_\mu$ and $V_G=V_\mu=0$~\cite{Moffat}. The first Friedmann equation now becomes
\begin{equation}
\label{Friedmann4}
H^2=\frac{8\pi G_N\rho_\phi}{3}+\frac{\Lambda}{3},
\end{equation}
where from (\ref{Vphi}) and (\ref{Tphi}) assuming that $W(\phi)\sim 0$:
\begin{equation}
\rho_\phi=\frac{1}{2}\omega\mu^2\phi_0^2.
\end{equation}

The equation for the density perturbation for the non-relativistic single-component fluid is given by the Jeans equation:
\begin{equation}
\label{deltaeq}
\ddot\delta_{\bf k}+2H\dot\delta_{\bf k}+\bigg(\frac{c_s^2a_0^2k^2}{a^2}-4\pi G_N\bar\rho\biggr)\delta_{\bf k}=0,
\end{equation}
where $\bar\rho$ denotes the mean density. Here, $c_s$ is the speed of sound:
\begin{equation}
c_s=\sqrt{\frac{dp}{d\rho}},
\end{equation}
and perturbations with $\delta p/\delta\rho=(d\bar {p}/dt)/(d\bar{\rho}/dt)=c_s^2$ are called adiabatic perturbations and for these perturbations $p=p(\rho)$. The first term in the parenthesis of (\ref{deltaeq}) is due to the adiabatic perturbation pressure contribution $\delta p=c_s^2\delta\rho$.

The nature of the solutions to (\ref{deltaeq}) depends on the sign of the factor in parenthesis. Pressure attempts to resist compression, so when the pressure term dominates, we obtain an oscillatory solution comprising standing density (sound) waves. The second term is due to gravity and when this term dominates the perturbations grow. The Jeans wavenumber when the pressure and gravity terms are equal is given by
\begin{equation}
k_J=\frac{a}{a_0}\frac{\sqrt{4\pi G\bar\rho}}{c_s},
\end{equation}
corresponding to the wavelength $\lambda_J=2\pi/k_J$ (Jeans length) and for non-relativistic matter $k_J\gg {\cal H}=(a/a_0)H$ and $c_s\ll 1$ where ${\cal H}$ is the comoving Hubble scale.

The Proca vector field $\phi_\mu$ is a neutral massive spin 1 particle (massive photon). Because it does not couple to massless photons {\it it can be treated as almost pressureless}, so the pressure gradient term in the Jeans equation is absent and $c_s\sim 0$. The Jeans length is approximately zero and we have
\begin{equation}
\label{phipertequation}
\ddot\delta_{\bf k}+2H\dot\delta_{\bf k}-4\pi G_N\bar\rho\delta_{\bf k}=0,
\end{equation}
where $\bar\rho=\bar\rho_r+\bar\rho_\phi$. For the radiation dominant era $\rho_\phi\ll\rho_r$, but perturbations in radiation, $\delta\rho_r$, oscillate rapidly and are damped $\delta\rho_r\sim 0$, so they can be ignored compared to the $\delta\rho_\phi$ perturbations. The expansion law is that for a radiation dominated universe:
\begin{equation}
a(t)\propto t^{1/2},\quad H=\frac{1}{2t}=\sqrt{\frac{8\pi G_N}{3}\bar\rho}.
\end{equation}
Dividing (\ref{phipertequation}) by $H^2$ we get
\begin{equation}
\frac{\ddot\delta_{\bf k}}{H^2}+2\frac{\dot\delta_{\bf k}}{H}=\frac{3}{2}\frac{\bar\rho_\phi}{\bar\rho}\delta_{\bf k}.
\end{equation}
The right-hand-side can be dropped because $\bar\rho_\phi < \bar\rho$. The solution to this equation is given by
\begin{equation}
\delta_{\bf k}=a+b\ln t,
\end{equation}
where $a$ and $b$ are constants. We find that the perturbations due to the phion particle grow at most logarithmically during the radiation dominated era of the universe. The increased expansion rate due to the presence of the smooth radiation component slows down the growth of perturbations.

Observations indicate that during the period between horizon entry, radiation-matter equality at a redshift $1+z_{eq}\sim 2.4\times 10^4\Omega_\phi h^2$  and decoupling, $\rho_b < \rho_\phi$. Because the phion particles are almost pressureless and the speed of sound $c_s\sim 0$, the phion Jeans length is approximately zero, so all scales are larger than the Jeans scale, and there is no oscillatory behavior for the phion particles. Instead, perturbations grow at all length scales. On the other hand, the $\rho_{b\gamma}$ perturbations oscillate before decoupling. The dark energy determined by the cosmological constant $\Lambda$ is a smooth component of energy and only becomes important at later times.

The baryon Jeans length is $\lambda_J=2\pi k_J^{-1}$, where
\begin{equation}
k_J^{-1}=\frac{a_0c_s}{a\sqrt{4\pi G_N\bar\rho}},
\end{equation}
and where now $c_s$ is the speed of sound for baryons. The ratio of the comoving Jeans length:
\begin{equation}
\lambda_J=\frac{2\pi a_0c_s}{a\sqrt{4\pi G_N\bar\rho}}
\end{equation}
to the comoving Hubble length:
\begin{equation}
{\cal H}^{-1}=\frac{a_0}{a\sqrt{8\pi G_N\bar\rho/3}}
\end{equation}
is given by
\begin{equation}
\label{Ratio}
{\cal H}\lambda_j=2\pi\sqrt{\frac{2c_s}{3}}.
\end{equation}
We see that $G$ has canceled in (\ref{Ratio}), so that the size of the effective $G$ in MOG cannot influence the rate of growth structure in the absence of the density $\rho_\phi$ or dark matter. However, we have that at the time of horizon entry and radiation-matter equality $G\sim G_N$.

Before decoupling the baryon Jeans length is comparable to the Hubble length, so adiabatic baryon perturbations will oscillate before decoupling. The baryon density perturbation begins to grow only after the decoupling time $t=t_{dec}$, because before decoupling the baryon-photon pressure prevents any growth. Without the phion particle density $\rho_\phi$ the baryon perturbations grow as $\delta_b\propto a\propto t^{2/3}$ after decoupling. However, the CMB anisotropy due to baryon density perturbations observed at $t=t_{dec}$ are too small $\sim 10^{-4}$ to produce a growth factor of 1100 needed to generate the presently observed large-scale structure. As in the standard CDM model, the phion particle density perturbations solve the problem, for they begin to grow earlier at horizon entry and by $t=t_{dec}$ they are significantly greater than the baryon density perturbations. After decoupling the baryon density perturbations $\delta\rho_b$ catch up to the phion particle density perturbations $\delta\rho_\phi$. The evolution behavior of the baryon and phion density perturbations is shown in Fig.1.\\

\includegraphics[scale=0.25]{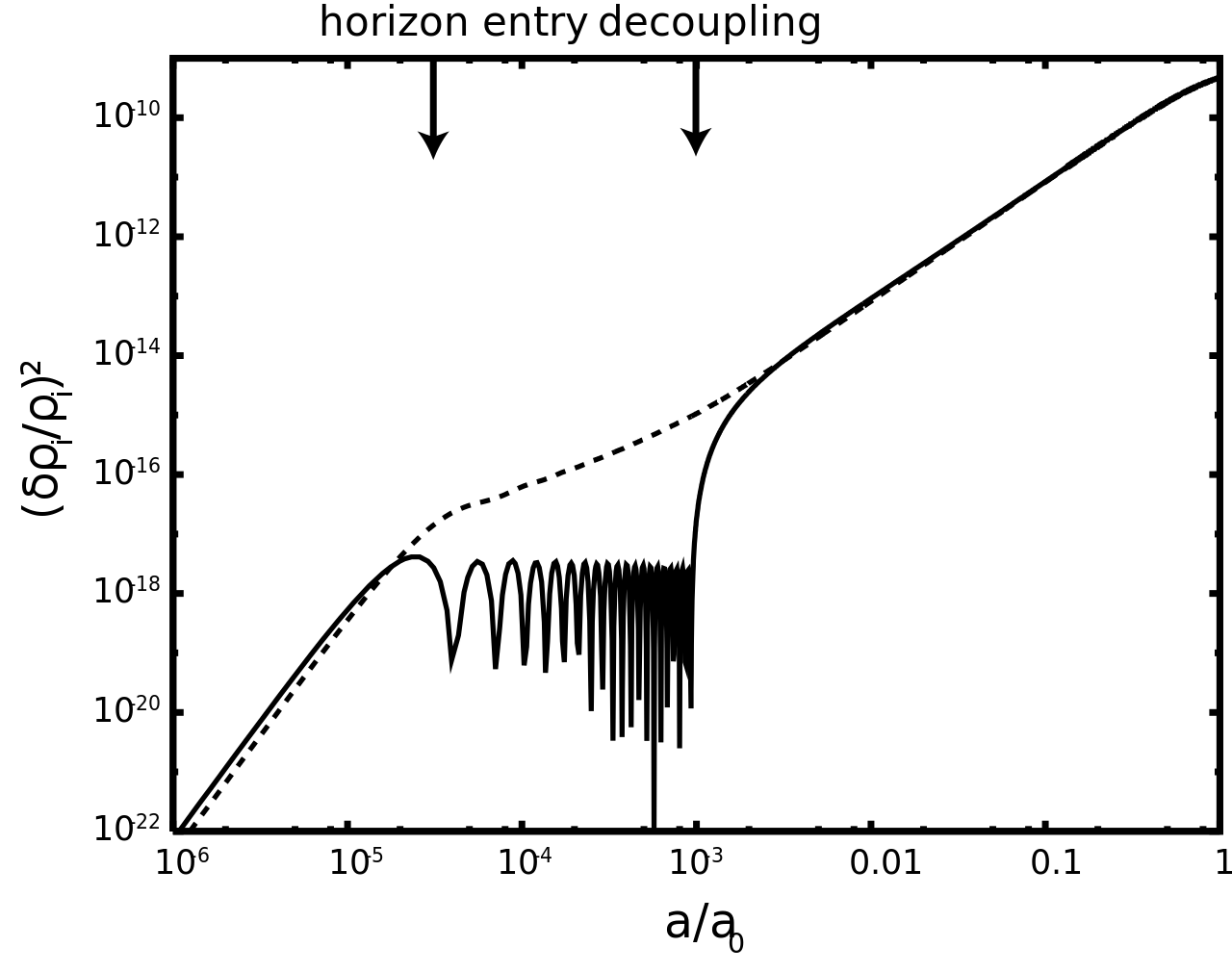}\\

Fig.1. Evolution of the phion matter and baryon density perturbations after horizon entry.  The baryon perturbations oscillate (solid black line) while the phion density perturbations (dashed line) grow after horizon entry.  The baryon perturbations catch up to the phion perturbations after decoupling.\\

As the universe expands beyond the time of decoupling, the gravitational attraction between the baryons becomes enhanced as the parameter $\alpha$ increases in size and $G=G_\infty=G_N(1+\alpha)$.  The density $\rho_\phi$ and the increasing enhancement of the size of $G$ deepen the baryon gravitational potential well. Eventually, as the large scale structures form $\rho_\phi < \rho_b$ and the baryon dominated MOG dynamics takes over. For the predictions of galaxy rotation curves and the galaxy cluster dynamics the best-fit values of $\alpha$ and $\mu$ are $\alpha=8.89\pm 0.34$ and $\mu=0.04\pm 0.004\,{\rm kpc}^{-1}$~\cite{MoffatRahvar1,MoffatRahvar2}.

The phion mass parameter $\mu$ is a variable scalar field and can evolve in size as the universe expands. After horizon entry and during the era approaching decoupling, the phion mass is $m_\phi\gg 10^{-28}\,{\rm eV}$ corresponding to cold dark matter. The phion mass could be as large as $10^{-6} < m_\phi < 1\,{\rm eV}$. When the earliest stars and galaxies form at about 400 million years after the big bang, $\mu$ undergoes a significant decrease, and from the best-fit values from galaxy rotation curves and cluster dynamics~\cite{MoffatRahvar1,MoffatRahvar2}, $\mu=0.04\,{\rm kpc}^{-1}$, corresponding to $m_\phi=2.6\times 10^{-28}\,{\rm eV}$. The Compton wavelength, in natural units, $\lambda_c\sim 1/m_\phi$, of this ultra-light phion is of the size of galaxies. In our MOG scenario, the phion particle evolves in the early universe until the formation of large scale structure as moderately heavy cold dark matter.  After the large scale structures have formed, the phion becomes ultra-light hot dark matter, which does not contribute to the dynamics of stars, galaxies and clusters of galaxies. Future microwave cavity experiments could detect an utra-light hidden photon~\cite{Mardon}, although the mass $m_\phi=2.6\times 10^{-28}\,{\rm eV}$ may be too small to detect. The mass and temperature evolution of the phion massive photon is the opposite of the standard axion-like dark matter particle. In the latter scenario, the axion is a hot dark matter particle in the early universe contributing to the density of radiation before decoupling, and after decoupling becomes a non-relativistic cold dark matter particle that can form stable galaxy and cluster halos.

\section{Power Spectra and the CMB}

The power spectrum as a function of wave number can be calculated from a transfer function $T(k)$. If the initial power spectrum and the transfer function are known, then the power spectrum at a later time can be calculated as
\begin{equation}
P(k)=T^2(k)P_0(k),
\end{equation}
where $P(k)$ is a dimensioned quantity. A dimensionless form of the power spectrum is
\begin{equation}
\Delta^2=Ak^3T^2(k)P_0(k),
\end{equation}
where $A$ is a normalization constant determined by observation. The initial power spectrum is believed to be almost scale invariant:
\begin{equation}
P_0(k)\propto k^n,
\end{equation}
where from the Planck measurement $n=0.9616\pm 0.0094$~\cite{Planck}.

In our MOG scenario the transfer function can be written as
\begin{equation}
T(k)=T_b(k)+T_\phi(k),
\end{equation}
where $T_b$ and $T_\phi$ denote the baryon and phion particle contributions, respectively. The baryon sound wave oscillations due to the baryon-photon pressure prior to the decoupling time produce acoustical peaks in the angular power spectrum, ${\cal D}_\ell=\ell(\ell+1)C_\ell/2\pi$. Because the form of the MOG Friedmann equation (\ref{Friedmann3}) is the same as in the ${\Lambda}CDM$ model, we can match the $\Lambda$CDM calculation of the CMB angular power spectrum. In the present day universe $\rho_b$ has the same value as in the ${\Lambda}CDM$ model, and $\Omega_b=8\pi G_N\rho_b/3H^2$ matches its value in the ${\Lambda}CDM$ model, using Newton's constant (non-gravitational effect).  Moreover, today $\rho_\phi\ll\rho_b$ and $G=G_N(1+\alpha)$, so for a constant non-zero value of $\alpha$, we have $(G_N\rho)_{\Lambda CDM}=(G_N(1+\alpha)\rho)_{\rm MOG}$, where $\rho_{\Lambda CDM}=\rho_b+\rho_{CDM}$ (plus small contributions from neutrino density) and $\rho_{\rm MOG}=\rho_b$. Now, redshifting towards the CMB and decoupling times, $\rho_\phi$ becomes smoothly bigger than $\rho_b$ and $(G_N\rho)_{\rm \Lambda CDM}=(G_N\rho_\phi)_{\rm MOG}$ as $G\rightarrow G_N$ and $\alpha$ becomes less than unity. It follows that the calculation of the CMB power spectrum in the ${\Lambda}CDM$ model can be duplicated in MOG, using the Planck 2013 best-fit values~\cite{Planck}: $\Omega_bh^2=0.022199$, $\Omega_ch^2=0.11847$, $n_s=0.9624$, $H_0=67.94\,{\rm km}\,{\rm sec}^{-1}\,{\rm Mpc}^{-1}$, together with the remaining parameters in the fitting process. The value of $\alpha$ today is $\alpha=4.3367$.\\

\includegraphics[scale=0.5]{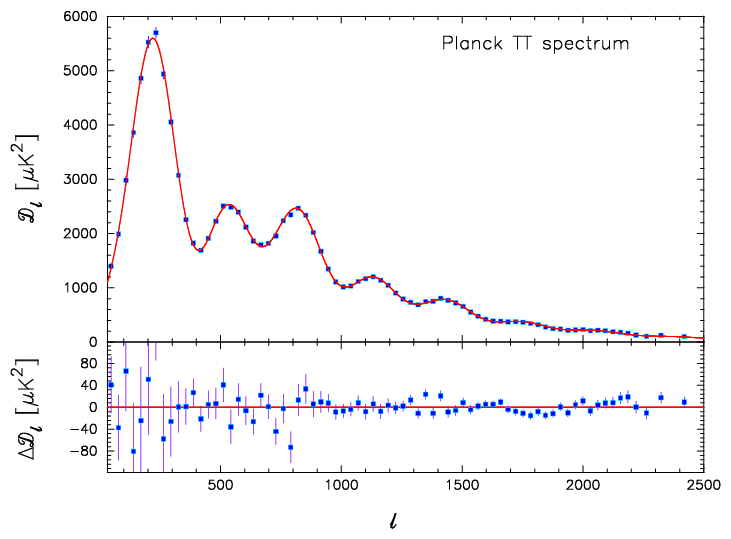}\\

Fig. 2. The TT power spectrum obtained from ref.~\cite{Planck}, using the best-fit parameters of the Planck+WP+highL fit listed in Table 5 of the Planck 2013 paper. The red line shows the best-fit MOG spectrum. The lower panel shows the residuals with respect to the MOG model (and the $\Lambda$CDM model).\\

Because for galaxies and clusters $\rho_\phi\ll \rho_b$ there is no significant halo of phion particles. Once the combined $\rho_b$ and $\rho_\phi$ density perturbations have grown sufficiently to produce stars and galaxies, then the MOG non-relativistic dynamics for baryons takes over to determine the final evolution and dynamics of galaxies. The transfer function $T_b$ for baryons has unit oscillations, which will show up in the calculation of the matter power spectrum. However, the finite size of galaxy survey samples and the associated window function used to produce presently available power spectra mask any such oscillations. Applying a window function to the MOG prediction for the matter power spectrum, smooths out the power spectrum curve. The enhanced size of $G=G_N(1+\alpha)$ with $\alpha$ non-zero predicts the right shape for the power spectrum curve, resulting in a fit to the data. The GR prediction without dark matter and with $G=G_N$ cannot produce the correct magnitude or shape for the matter power spectrum. In Fig. 3 we show the predicted MOG matter power spectrum~\cite{MoffatToth3}. This result was developed without the benefit of an extended and continuous baryon matter distribution. In future galaxy surveys which utilize a large enough number of galaxies, with galaxies detected at sufficiently large redshift $z$, and with the use of a sufficiently narrow enough window function, it should be possible to detect any significant oscillations in the matter power spectrum.

\includegraphics[scale=0.75]{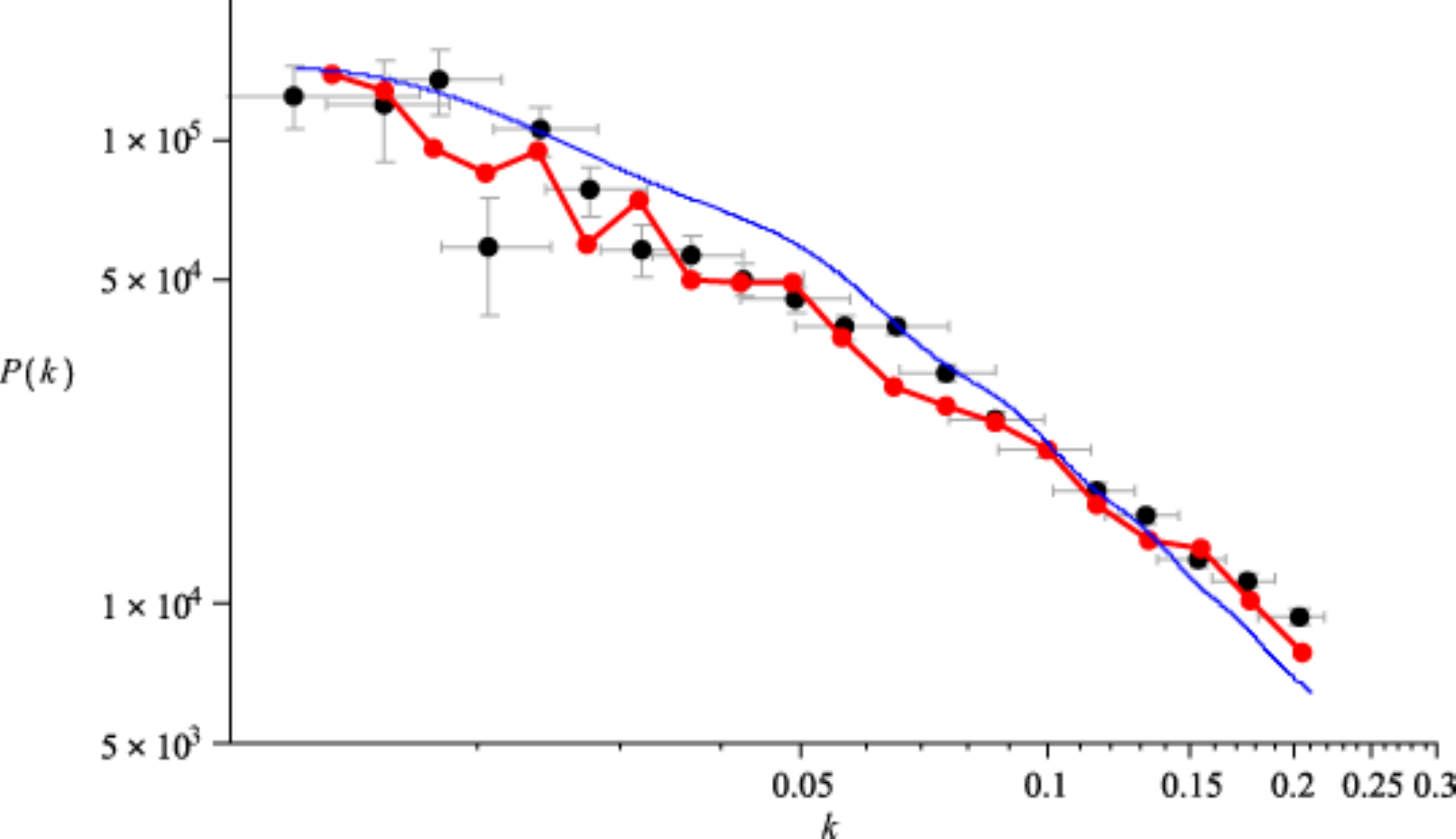}\\

Fig.3. After applying the appropriate window function, MOG (thick red line) shows agreement with the luminous red galaxy survey mass power spectrum data. The MOG fit to the data is comparable to the CDM prediction (thin blue line)~\cite{MoffatToth3}.

\section{Conclusions}

By assuming that the density of phion particles associated with the massive and neutral vector field $\phi_\mu$ in MOG theory is the dominant density in the early universe with a phion mass, $m_\phi\gg 10^{-28}\,{\rm eV}$, the perturbations $\delta\rho_\phi$ satisfy the pressureless Jeans equation, allowing for an enhanced growth from the time of horizon entry and radiation-matter equality to produce large scale stellar and galaxy structure after the time of decoupling. The baryons obey the Jeans equation with baryon-photon pressure, so that the baryons oscillate between the time of horizon entry and decoupling, generating baryon-pressure acoustical waves detected in the angular power spectrum in the CMB. After the universe expands beyond the decoupling time, the pion particle density and baryon density are comparable until the time of stellar and galaxy formation when $\rho_\phi\ll \rho_b$ and the mass of the phion becomes ultra-relativistic with $m_\phi=2.6\times 10^{-28}\,{\rm eV}$.  Then, the non-relativistic STVG Newtonian acceleration law, including the repulsive phion particle Yukawa interaction and the attractive gravitation due to the enhanced value of $G$, determines the rotation curves of galaxies and the dynamics of galaxy clusters. 

The angular power spectrum describing the imprint of the baryon-photon acoustical waves on the CMB is determined starting with $\Omega_\phi\ll \Omega_b$ today and $(G_N\rho)_{\Lambda CDM}=(G_N(1+\alpha)\rho)_{\rm MOG}$. Then, redshift scaling the calculation of the transfer function to the time prior to decoupling and the surface of last scattering 
with $(G_N\rho)_{\Lambda CDM}=(G_N\rho_\phi)_{\rm MOG}$, the angular power spectrum fit matches the $\Lambda$CDM model fit to the Planck 2013 data. The matter power spectrum in MOG can, with an appropriate window function, fit the galaxy matter distribution data. A critical test of MOG is whether significant baryon oscillations in the power spectrum begin to show as the number of observed large red galaxies increases and the size of the window function decreases. If the smooth ${\Lambda}CDM$ model fit to the matter power spectrum persists with a large enough increase in observed galaxies in galaxy redshift surveys, then this would rule out MOG in favor of the existence of dark matter halos in galaxies.

The ultra-light phion particle (hidden light photon) and its associated Proca field $\phi_\mu$ are an integral part of the modified gravitational theory determined by the STVG action. Because the MOG predicts observations from the solar system to galaxy clusters and the large scale structure of the universe, it can provide a fundamental description of the universe at both small and large distance scales. 

\section*{Acknowledgments}

I thank Niayesh Afshordi, Gordan Krnjaic and Viktor Toth for helpful discussions. This research was generously supported by the John Templeton Foundation. Research at the Perimeter Institute for Theoretical Physics is supported by the Government of Canada through industry Canada and by the Province of Ontario through the Ministry of Research and Innovation (MRI).


\begin{thebibliography}{10}

\bibitem{Wimp} Snowmass CF1 Summary: Cushman, P., et al. arXiv/1310.8327 [hep-ex]

\bibitem{Xenon} XENON100 collaboration: Aprile, E., et al. 2013, Phys. Rev. Lett, 111, 021301 

\bibitem{Lux} LUX Collaboration: Akerib, D.~S., et al. 2014, Phys. Rev. Lett. 112, 091303 

\bibitem{Moffat} Moffat, J.~W.\ 2006, JCAP 0603  004

\bibitem{BrownsteinMoffat1} Brownstein, J. ~R., \& Moffat, J.~W. 2006, 367, 527

\bibitem{BrownsteinMoffat2} Brownstein, J. ~R., \& Moffat, J. ~W. 2006, ApJ, 636, 721

\bibitem{BrownsteinMoffat3} Brownstein, J. ~R., \& Moffat, J. ~W. 2007, MNRAS 382, 29

\bibitem{Brownstein} Brownstein, J. ~R. 2009, Ph.D. Thesis, University of Waterloo.

\bibitem{MoffatToth1} Moffat, J. ~W., \& Toth, V. ~T., 2008, ApJ, 680, 1158

\bibitem{MoffatToth2} Moffat, J. ~W., \& Toth, V. ~T., 2009, MNRAS, 397, 1885

\bibitem{MoffatToth3} Moffat, J. ~W., \& Toth, V. ~T., 2009, Class. Quant. Grav. 26 (8), 085002

\bibitem{MoffatToth4} Moffat, J. ~W., \& Toth, V. ~T., 2013, Galaxies, 1 (1), 65

\bibitem{MoffatRahvar1} Moffat, J.~W., \& Rahvar, S. 2013, MNRAS, 436, 1439

\bibitem{MoffatRahvar2} Moffat, J.~W., \& Rahvar, S. 2014, MNRAS, 441, 3724

\bibitem{Roshan1} Roshan, M, 2013, Phys. Rev. D 87, 044005

\bibitem{Roshan2} Roshan, M., \& Abbassi, S. 2014, Phys. Rev. D 90, 044010

\bibitem{Planck} Planck Collaboration:  Ade, P. ~A. ~R. et al. arXiv:1303.5076v3

\bibitem{Mardon} For references to hidden photon theory and its experimental detection: Graham, P. ~W., et al. arXiv/1407.4806 [hep-ph]. 


\end{thebibliography}
\end{document}